\documentclass[aps,prb,groupedaddress,twocolumn,10pt]{revtex4-1}

\begin{document}

%\makeatletter
%\edef\orig@output{\the\output}
%\output{\setbox\@cclv\vbox{\unvbox\@cclv\vspace{0pt plus 5pt}}\orig@output}
%\makeatother

\newcommand{\be}{\begin{equation}}
\newcommand{\ee}{\end{equation}}
\newcommand{\bea}{\begin{eqnarray}}
\newcommand{\eea}{\end{eqnarray}}

\title{SYK does not Transit Gloria Mundi just yet}

\author{D. V. Khveshchenko} 
\affiliation{Department of Physics and Astronomy, 
University of North Carolina, Chapel Hill, NC 27599}

\begin{abstract}
\noindent
This note discusses examples of $0+1$-dimensional Liouvillean dynamics  
instigated by the various deformations of the Sachdev-Ye-Kitaev (SYK) model. 
In reference to such deformations the main focus is on the regions of parameter space where 
the competing SYK couplings are of comparable strength and can not be treated as each others  
perturbations in the vicinity of the conformal fixed points
corresponding to the pure $SYK_q$ models with different values of $q$.
Crossovers between such fixed points ('SYK transits') 
can be efficiently studied in the equivalent framework 
of single-particle quantum mechanics. 
\noindent
\end{abstract}
\maketitle

%\begin{frontmatter}

\title{SYK does not TRANSIT GLORIA MUNDI just yet}

\author{D.V.~Khveshchenko} 
\address{Department of Physics and Astronomy, University of North Carolina, Chapel Hill, NC 27599, U.\,S.\,A.
\email{khvesh@physics.unc.edu}
}

\received{2 June 2022}
\revised{ 2022}
\accepted{ 2022}
%\runningauthor{D.V.~Khveshchenko}

\section{The rise of SYK}

The glorious rise of the celebrated 
SYK model \cite{sy01,sy02,sy03,sy04,sy05,sy06,sy07,sy08} into one of the central themes in modern interdisciplinary 
theoretical studies 
was due to a rare confluence of such precious properties as its elegant solubility, maximally chaotic behavior, asymptotic conformal symmetry, and more.  
The numerous in-depth analyses of the SYK model revealed a number of important connections between such seemingly disjoint subjects  
as random matrices, quantum black holes, disordered quantum dots, and, possibly, strange metallic behaviors in the various condensed matter systems.  

One of these novel connections may have already contributed towards a resolution of the long-standing black hole information paradox by demonstrating that the properly (re)defined Hawking radiation entropy can be unitary, following the previously predicted Page curve \cite{bh01,bh02}.

In the condensed matter context, the SYK model has served as a powerful 
inspiration for a great many proposed non-Fermi-liquid (NFL) scenarios \cite{21a,21b,21c,21d,21e,21f,21g,21h,21i,21j,21k,21l,21m,21n,21o,21p}.  
However, the very existence of numerous  
plausible explanations of, e.g., the ubiquitous linear temperature 
dependence of resistivity in bad metals \cite{41a,41b,41c,41d,41e,41f,41g,41h,41i,41j,41k,41l,41m,41n,41o,41p,41q}
may seem to suggest that its ultimate explanation is yet to be found. 

Nevertheless, alongside the renewed interest in hydrodynamics inspired by the holographic ideas,  
the SYK scenaria have been particularly important for pursuing the ad hoc field of 'bottom-up' (a.k.a.'non-AdS/non-CFT') holography which purports
to describe a variety of (allegedly) strongly correlated condensed matter systems \cite{ads01,ads02,ads03,ads04,ads05,ads06,ads07}. Indeed, with the once abundant and defiantly upbeat claims of 'explaining' high-$T_c$ materials, heavy fermions, graphene, etc. by virtue of some uncontrolled calculations in the conveniently chosen (and/or previously studied) classical gravity theories all but gone, the SYK model has remained   
a rather unique theoretical playground for obtaining rigorous results. 

In that regard, the SYK model would be often referred to as a genuine example of low-dimensional
holographic correspondence - even despite the fact that, 
both being effectively one-dimensional, the 
low-energy sector of SYK and its dual (formally, two-dimensional) Jackiw-Teitelboim (JT) 
gravity present a form of equivalence between different  
realizations of the quantized co-adjoint orbit of the (chiral) Virasoro group.

Such equivalence does not quite 
rise to the level of full-fledged holographic duality, as the JT bulk dual is 
non-dynamical and determined by the boundary degrees of freedom. By contrast,  
in order to qualify as a true holographic scenario  
the bulk theory would have to have some non-trivial bulk dynamics 
that gets quenched and turns classical only in a certain
('large-$N$') limit \cite{ads01,ads02,ads03,ads04,ads05,ads06,ads07}. 

Moreover, similar remarks can also be made about the (historically, somewhat less extensively discussed) correspondence between the $3d$ gravity with BTZ-like black hole backgrounds  
and the various (KdV and alike) families of solvable $1+1$-dimensional systems  
(see, e.g., Ref.~\cite{dvk2} and references therein).\\

\section{The SYK deformations}

Since the beginning of the SYK era there have been attempts to explore deviations from the original $SYK_4$ 
model in order to assess the generality (or, conversely, uniqueness) of the behavior that it represents. In particular, there has been much discussion of the conjectured 
NFL-Fermi liquid 
(FL) transition in the hybrid $SYK_4-SYK_2$ model \cite{sy01,sy02,sy03,sy04,sy05,sy06,sy07,sy08,21a,21b,21c,21d,21e,21f,21g,21h,21i,21j,21k,21l,21m,21n,21o,21p}. 

A renormalization flow between the two fixed points has been mostly 
studied by means of perturbation theory 
operating in terms of the propagator $G(\tau_1,\tau_2)$
of $N\gg 1$ spaceless Majorana fermions \cite{21a,21b,21c,21d,21e,21f,21g,21h,21i,21j,21k,21l,21m,21n,21o,21p}. 
In the conformal limit of the generalized $SYK_q$ of order $q\geq 4$ the latter exhibits
the fermion dimension $\Delta=1/q$, thus making the perturbation proportional to $G^{q/2}$ strongly relevant (dimension one) in the near vicinity of the UV
$SYK_q$ fixed point.  
Conversely, the formerly leading term $G^q$ becomes strongly 
irrelevant (dimension four) near the IR fixed point of $SYK_{q/2}$.
A unique feature of the $q=4$ case, though, is that the transition
occurs not between two different NFLs but the $SYK_4$ NFL and the disordered FL.

Notably, in the course of crossing over between the different fixed point regimes   
the value of $q$ of the dominant term plays a role akin to that of the central charge
in $2d$ conformal field theories. 

As far as the potential physical applications are concerned, some
of the previous analyses \cite{21a,21b,21c,21d,21e,21f,21g,21h,21i,21j,21k,21l,21m,21n,21o,21p,7a,7b,JY,lunkin1} suggest that the putative phase transition may take place
at critical couplings vanishing as powers of  
$1/N$ - which value would be practically indistinguishable from zero in a macroscopic   
system - while others yield values that remain finite in the $N\to\infty$ limit.

The common approach to a SYK-type model
starts out by integrating the fermions out, thereby  
arriving at the action in terms of the bi-local fields $G(\tau_1,\tau_2)$ and the corresponding 
self-energy $\Sigma(\tau_1,\tau_2)$
\bea
S={N\over 2}Tr\ln(\partial_{\tau}-\Sigma)+\nonumber\\
{N\over 2}\int\int d{\tau_1}d{\tau_2}(\Sigma(\tau_1,\tau_2) G(\tau_1,\tau_2)-
F[G(\tau_1,\tau_2)])
\eea
where the functional $F[G]$ results from averaging over the  
Gaussian-correlated random amplitudes of all-to-all $q$-body entanglement.
Moreover, such entangling couplings 
can be made non-uniform, thus introducing a notion of spatial dimensions and further extending the class of 
attainable models to include those with 'distance'-dependent entanglement \cite{dvk1a,dvk1b}. 

Solving for the self-energy, the Schwinger-Dyson equation derived from (1) takes the form 
\be
\partial_{\tau_1}G(\tau_1,\tau_2)-\int d{\tau_3}<\tau_1|{\delta F\over \delta G}|\tau_3>G(\tau_3,\tau_2)\nonumber\\
=\delta(\tau_1-\tau_2)
\ee
In the original $SYK_q$ model with $F(G)=J^2G^q$ 
the equation (2) remains invariant under any
diffeomorphisms $\tau\to f(\tau)$ of the thermodynamic time variable $\tau$ 
subject to the boundary condition $f(\tau+\beta)=f(\tau)+\beta$ as long as
the derivative term is neglected and provided that $G$ and $\Sigma$ transform as 
\bea 
G(\tau_1,\tau_2)\to G_f=(f^{\prime}(\tau_1)f^{\prime}(\tau_2))^\Delta G(f(\tau_1),f(\tau_2))~~~\\
\Sigma(\tau_1,\tau_2)\to \Sigma_f=(f^{\prime}(\tau_1)f^{\prime}(\tau_2))^{1-\Delta}\Sigma
(f(\tau_1),f(\tau_2))
\nonumber
\eea
At $J\tau\gg 1$ a representative power-law solution
to the equation (2) 
reads $G_0(\tau_1,\tau_2)\sim sgn\tau/(J\tau)^{2\Delta}$ (hereafter $\tau=\tau_1-\tau_2$). 

Choosing a particular 
mean-field solution reduces the invariance under arbitrary diffeomorphisms 
down to the subgroup of the Mobius transformations  $SL(2,R)$.
Correspondingly, a gradient expansion of the logarithm in (1) yields the 
(approximately) local effective action which describes 
finite temperature dynamics of the reparametrization mode \cite{syk01,syk02,syk03,syk04,syk05,syk06,syk07,syk08,syk09,syk10,syk11,syk12,syk13,syk14,syk15,syk16,syk17,syk18,syk19,syk20,syk21,syk22} 
\be
S_0(f)=-O(1){N\over Jq^2}\int Sch{\{}\tan{\pi f\over \beta},\tau {\}}d\tau
\ee
where $Sch$ stands for the Schwarzian derivative 
%\be
$Sch{\{}f,x {\}}= {f^{\prime\prime\prime}\over f^{\prime}}-{3\over 2}
({f^{\prime\prime}\over f^{\prime}})^2
$
%\ee
obeying the differential 'chain rule'
%\be
$Sch {\{}F(f),x{\}}=Sch {\{}F(f),f{\}}{f^\prime}^2+Sch {\{}f,x{\}}$
%\ee
and operating on the manifold of (nearly)  
degenerate states related by virtue of the transformations (3).\\

\section{Liouvillean quantum mechanics}

Under the customary parametrization $f^{\prime}(\tau)=e^{\phi(\tau)}$
the Schwarzian action (4) assumes the (pseudo-)free form
%\be
$S_0(\phi)\sim \int d{\tau} (\phi^{\prime})^2$
%\ee
.
In the process of averaging the products of propagators
\bea 
<G_f(\tau_1,\tau_2)\dots G_f(\tau_{2p-1},\tau_{2p})>=\nonumber\\
=\int D\phi e^{-S_0(\phi)}
\prod_{i=1}^p
{e^{\Delta(\phi(\tau_{2i-1})+\phi(\tau_{2i}))}\over (\int_{\tau_{2i-1}}^{\tau_{2i}}d\tau e^{\phi})^{2\Delta} }
\eea
over the fluctuations of $\phi$ the action $S_0(\phi)$ gets augmented 
by the Liouville term 
%\be
$\Delta S_2(\phi)=h_2\int d{\tau}e^{2\phi(\tau)}$
%\ee 
with $h_2\sim J$.
Technically, upon promoting the denominator  in (5) to the exponent 
with the help of some auxiliary  integration a la Feynman the overall effective potential 
acquires a piece-wise  Liouville term 
acting during 
the time intervals between $2p$ consecutive insertions of the 
operator $e^{i\Delta\phi}$ \cite{bak01,bak02}. 

The resulting action $S_0+\Delta S_2$ can then be quantized by considering 
the corresponding (rescaled) Schroedinger equation \cite{bak01,bak02}
\be
(-{1\over 2}\partial_{\phi}^2+h_2e^{2\phi})\psi=E\psi
\ee 
whose scattering states 
$\psi_k(z)\sim K_{2ik}(2{\sqrt z})$ (here $z=\lambda e^{\phi}$)
belong to the continuum with the spectrum $E_k=k^2$ and density of states
$\rho(E)\sim\sinh{2\pi{\sqrt {2NE/J}}}$ \cite{syk01,syk02,syk03,syk04,syk05,syk06,syk07,syk08,syk09,syk10,syk11,syk12,syk13,syk14,syk15,syk16,syk17,syk18,syk19,syk20,syk21,syk22,bak01,bak02}.

These exact expressions can be used to compute the matrix elements 
$<0|e^{\Delta\phi}|k>$ exactly. Such calculations 
reveal the universal limit of an arbitrary power of $G_f$ averaged over the soft mode fluctuations in the late-time, 
$\tau > N/J$ -  
or, at finite temperatures, in the strong coupling,
$J\beta/N\gtrsim 1$ - regime \cite{bak01,bak02},  
\be
<G^p_f>\propto {1\over (J\tau)^{3/2}}
\ee
This behavior is markedly different from the (non-universal)  mean-field one 
$
G^p_0\propto 1/{\tau}^{2p/q}
$
at shorter times ($\tau > N/J$) or weak couplings ($J\beta/N\lesssim 1$).

In that regard, the presence of the exponential term $\Delta S_2(\phi)$ in the overall 
effective action is instrumental. In its absence the Gaussian 
fluctuations of the field $\phi$ governed by $S_0(\phi)$ would  
have caused non-algebraic decay, thus 
being unable to deliver the universal power-law (7).
In fact, such a behavior could have never emerged  
out of the purely Gaussian $\phi$ fluctuations even if 
the correlator $<\phi(\tau)\phi(0)>$ were logarithmic, as
 $\ln <G^p_f>$ would still depend on $p$ and $\Delta$ (in both cases, quadratically).

The effective action might also 
include the various intrinsically non-local terms 
\be
\Delta S_n={h_n}
\int\int d\tau_1d\tau_2
({f^{\prime}(\tau_1)f^{\prime}(\tau_2)\over (f(\tau_1)-f(\tau_2))^2})^{\Delta_n}
\ee
which can dominate over (4) for $\Delta_n\leq 3/2$. In the previous analyses, such terms would be 
routinely substituted with the local operators
$
\Delta S_n=h_n \int d{\tau} e^{\Delta_n\phi(\tau)}
$
thus further modifying the equivalent quantum mechanical Hamiltonian in (6).\\

\section{Bi-quadratic SYK deformation}

This important example of the deformed SYK model 
has been extensively discussed  in the context of random tunneling between two different SYK systems. For example, it arises in such, at first sight, 
unrelated fields as theoretical 
cosmology ('traversable wormhole') \cite{syk01,syk02,syk03,syk04,syk05,syk06,syk07,syk08,syk09,syk10,syk11,syk12,syk13,syk14,syk15,syk16,syk17,syk18,syk19,syk20,syk21,syk22} and coupled quantum dots \cite{dot01,dot02,dot03,dot04,dot05,dot06,dot07,dot08,dot09,dot10,dot11}.  

In most analyses, the perturbed propagator would be taken in the form (3) of a 'gauge-transformed' mean-field solution $G_0$, 
thereby accounting for the 'soft' reparametrization mode $f(\tau)$ 
while ignoring any potential changes to the mean-field background field configuration
itself.

In particular, adding the  $SYK_2$ ('tunneling') term with the amplitude $\Gamma$ 
replaces the Liouville potential in the $SYK_4$ action  
(written in the Euclidean signature) with the Morse-type one \cite{MQ01,MQ02,MQ03,MQ04} 
\be
S=\int d{\tau}
({1\over 2J}({\phi}^{\prime})^2+Je^{2\phi} + {\Gamma^2\over J} e^{\phi}+{1\over \beta^2J}e^{4\phi})
\ee  
The corresponding Schroedinger equation with the (properly rescaled) Hamiltonian 
\be
H=-{1\over 2}\partial_{\phi}^2+e^{2\phi}+\lambda e^{\phi}+{\lambda^{\prime}}e^{4\phi}
\ee 
which can be solved exactly 
in terms of the wave functions (here $z=2\lambda e^{\phi}$) 
%\be
$\psi_{k}(z)\sim e^{-\phi/2} W_{\lambda, ik}(z)$
%\ee
with the continuous spectrum $E_k=k^2+1/4+\lambda^2$ if the last - subdominant at low temperatures (or large negative $\phi$) - term in (10) can be dropped  \cite{MQ01,MQ02,MQ03,MQ04,lunkin2}.  

Besides, for $\lambda<0$ the Hamiltonian (10) appears to possess a finite number of bound states 
\be
\psi_{n}(z)\sim
z^{\lambda-n-1/2-z/2} L_n^{2\lambda-2n-1}(z)
\ee
at the discrete energies  
$E_n=-(\lambda-n+1/2)^2$, $n=0,\dots, [\lambda-1/2]$.
Near its minimum this spectrum  
can be approximated by the oscillator one. 

A somewhat different path leading up to the Morse-type action (9) was taken in Ref.~\cite{lunkin2}. 
The effect of the tunneling term with $\Delta_1=1/2$ was argued to be two-fold:
first, it contributes to (and/or refines) the 
purely Schwarzian (or 'hard' mode) saddle-point solution 
and, second, controls the pseudo-Goldstone (or 'soft' mode) fluctuations. 
These roles would be separately played by the 'longitudinal' 
(or radial, $e^{\xi}=1-f^{\prime}$, in the holographically dual JT picture)  
and 'transverse' (or angular, $\phi$) fluctuations, respectively. 
The former were argued to be strongly non-Gaussian and the 
effect of such fluctuations was claimed in Ref.~\cite{lunkin2} to strengthen (somewhat unexpectedly) 
 the $SYK_4$ conformal mean-field behavior over a broader range of parameters. 

More specifically, the strong coupling Schwarzian regime 
was argued to sustain the $SYK_2$
perturbation at all couplings $\gamma\equiv\Gamma/J$ below $\gamma_c\sim 1/N$
while at its higher values the propagator was 
found to crossover to the $q=2$ FL fixed point.
This was argued to be suggestive of a
zero-temperature phase transition taking place at $\gamma_c$, rather than at a much larger 
value of order $1/N^{1/2}$, as per the naive estimate. 
Such parametric reduction of $\gamma_c$ was claimed to manifest a stabilizing 
effect of the $SYK_2$ coupling on the mean-field conformal solution 
against the Schwarzian fluctuations due to the formation 
of a polaron-like non-perturbative field configuration. 

Correspondingly, the earlier perturbative analysis by the same authors
revealed that a weak $SYK_2$ coupling does not alter the Schwarzian 
asymptotic (7) up to the values of $\gamma$ of order $\gamma_c$ \cite{lunkin1}.   

Such observations appear to be generally consistent with those of Refs.~\cite{35a,35b}
which conjectured the existence of a chaotic-integrable transition in the $SYK_q-SYK_2$ model  
at finite temperatures. 
Above the transition temperature the system was found to behave chaotically while 
below it the chaos-related Lyapunov exponent (see below) dropped to zero, thus hinting at  
the FL nature of the underlying ground state.

On the technical side,  
upon first introducing two Lagrange multipliers $\lambda$ and $\Lambda$ 
and then voluntarily relaxing the corresponding constraints 
by fixing their mean-field values, Ref.~\cite{lunkin2} arrived at the effective action
\bea
S=\int d{\tau}
({1\over 2}(\phi^{\prime})^2+\Lambda(e^{2\phi} - f^{\prime})+\lambda (e^{\phi}-\chi)\nonumber\\
-{1\over (\beta J)^2}e^{4\phi})-{1\over 2}\int\int d{\tau}_1d{\tau_2}
{\chi(\tau)\chi(\tau^{\prime})\over {|f(\tau_1)-f(\tau_2)|} }
\eea  
In this (perhaps, somewhat excessive) parametrization, functional 
integration about the mean-field $SYK_4$ 
fixed point factorizes into first taking a quantum mechanical expectation value over the exact ground state
$\psi_0(\xi)$ of the Hamiltonian (10)
and then additionally averaging over the Gaussian (perturbative)  
$\phi$ fluctuations. In Ref~\cite{lunkin2} neither mechanism was 
found to have any profound effect on the correlators, though. 

In particular, an arbitrary power of 
the mean-field propagator would still retain its bare mean-field form 
provided that the $\phi$-fluctuations were controlled by a large parameter $\lambda$.
Likewise, averaging over the ground state of (10) adds the square of
a non-singular expectation value
$
<0|e^{p\Delta\phi}|0> = \int d\phi e^{p\Delta\phi}\psi_0^2(\phi)
$
which does not give rise to any decaying power-law factor either.
In that sense, the largely negligible effect of, both, the Gaussian fluctuations and the ground state averaging may indeed be viewed as increased stability of the mean-field regime 
in the presence of even a small $SYK_2$ coupling. 

It should be noted, though, that under the assumption of $\lambda<0$  
the Morse potential in (10) appears to differ from that of Refs.\cite{7a,7b,JY}
which is strictly repulsive, monotonic ($\lambda>0$), and lacks any bound states. 
It might also be concerning that if the potential in (10) were to 
support any bound states with $E_n<0$ then 
the fluctuation-averaged two-point correlator 
$
<G_f(\tau)>=\sum_{n}e^{-E_n\tau}N(E_n)
$
would be receiving  
- on top of the universal term (7) that stems from the continuum of  
scattering states with $E_k>0$ - a non-unitary (exponential) contribution 
whose potential divergence could only be arrested 
by the squared matrix element $N(E_n<0)=|<0|e^{\Delta\phi}|n>|^2$.
 
Interestingly, for $J=\Gamma$ the 
aforementioned monotonic and non-monotonic Morse  potentials
represent two super-partners fitting into one super-symmetric pair
%\be
$W_{\pm}(\phi)=V^2\pm{dV/d\phi}$
%\ee
with $V(\phi)\propto e^{\phi}$.    
The ground state of the binding potential  
then takes the form $\psi_0(\phi)\propto\exp(-\int Vd\phi)$.

Conceivably, the effective action $S(\phi)$ 
may develop other interesting regimes at the points of still higher symmetry.    
One such example would be provided by the Hulten potential
\be 
W(\phi)=\lambda {e^{\phi}\over 1-e^{\phi}}
\ee
whose first three terms of the expansion in powers of $e^{\phi}$
coincide with the 'hyper-symmetric' (or 'tri-critical') point 
$J=\Gamma=1/\beta$ in (9).
On the other hand, the $1/\phi$-behavior at small negative $\phi$ would be 
similar to that in the Coulomb potential,  
although the potential (13) features only a finite number ($[\lambda]$) of bound 
states at $E_n=-(\lambda^2-n^2/2\lambda n)^2$.
\\

\section{Large $q$ limit}

An alternate approach to the SYK models exploits
the large-$q$ approximation where the propagator is sought out in the form 
\be
G(\tau)={1\over 2}sgn \tau (1+{2\over q}g(\tau)+\dots)
\ee
Higher order corrections in $1/q$ can also be evaluated, albeit at the
increasingly prohibitive costs  \cite{syk01,syk02,syk03,syk04,syk05,syk06,syk07,syk08,syk09,syk10,syk11,syk12,syk13,syk14,syk15,syk16,syk17,syk18,syk19,syk20,syk21,syk22}.  

In action in the path integral over the field $g$
then takes the form 
\be
S(g)={N\over q^2}
\int\int d\tau_1d\tau_2 ({1\over 2}{dg\over d\tau_1}{dg\over d\tau_2}+W(g))
\ee
with the corresponding equation of motion 
\be
\partial^2_{\tau}g=-{\partial W(g)\over \partial g}
\ee
Formally solving (16) one obtains the classical trajectory
\be
\tau=\int^{0}_{g_0} {dg\over {\sqrt {W(g_0)-W(g)}}}
\ee
with the use of which thermodynamics of the system can be studied 
by putting $\tau=\beta/2$ \cite{7a,7b,JY}.
In particular, the turning point $g_0<0$ of the potential 
can be directly related to the mean-field energy \cite{JY} 
\be
E={N\over 4q^2}(\beta W(g_0)-2^{3/2}\int^{0}_{g_0} {dg{\sqrt {W_0-W(g)}}})
\ee
As already mentioned, one possible generalization of the bi-quadratic  
(Schwarzian plus tunneling) $q=4$ action to the larger values of $q$ 
is provided by the $SYK_q-SYK_{q/2}$ functional   
\bea
F[G]={2^qJ^2\over q^2}G^q(\tau_1,\tau_2)+{2^{q/2}\Gamma^2\over q^2} G^{q/2}(\tau_{1},\tau_2)
\eea
The corresponding effective potential
\be
W(g)={J^2}e^{2g}+{\Gamma^2}e^{g}
\ee 
allows for the explicit saddle point solution \cite{7a,7b,JY} 
\be
g(\tau)=-\ln(1+{\sqrt {J^2+4\Gamma^2}}\tau+\Gamma^2\tau^2)
\ee 
that gives rise to the mean-field propagator 
\be
G_0(\tau)={1\over 2}{sgn \tau\over (1+{\sqrt {J^2+4\Gamma^2}}\tau+\Gamma^2\tau^2)^{2/q}}
\ee
For future reference, the final-temperature counterpart of (21) reads
$g(\tau)=-\ln[({\sqrt {v^2J^2/\beta^2+\Gamma^4}}\cos(2v\tau/\beta-v)+\Gamma^2)(\beta^2/2v^2)]$
where the parameter  $v$ is to be determined from the relation
$2v^2=\Gamma^2\beta^2+\cos v{\sqrt {J^2v^2\beta^2+\Gamma^4\beta^4}}$ and becomes 
$v=1-O(1/\beta J)$ for $\Gamma=0$ \cite{7a,7b,JY}. 

It is worth noting that, 
in the look-alike equations (10) and (20) the field variables $\phi$ and $g$ depend on 
the 'center-of-mass' (cf. Eq.(8)) and relative times, respectively.
Also, unlike the approximate conformal propagator $G_0$, the expression (22) 
is UV-finite and naturally regularized at $\tau\sim min[1/J,1/\Gamma]$.
Hence, by contrast with the latter,
the saddle-point solution (22) remains applicable 
at all $\gamma$, both large and small. 
Therefore, the fluctuations of $g(\tau)$ describe pseudo-Goldstone excitations 
about the fixed 'valley'
in the space of field configurations which no longer needs to be adjusted.
\\  

\section{Quadratic fluctuations}

Small fluctuations about the mean-field solution (22)
are described by the Gaussian action  
\be
S_2={N\over 2q^2} \int\int d\tau_1d{\tau_2} 
\delta g(\tau_1)
{\partial^2S\over \partial g^2}|_{g_0}\delta g(\tau_2)
\ee
For a potential $W(g)=\sum_nc_ne^{ng}$ these fluctuations $\delta g$ would then be 
governed by a functionally similar kernel ${\partial^2W/\partial g^2}=\sum_nc_nn(n-1)e^{ng}$.
Albeit similar in its appearance to the previously discussed $S(\phi)$, this action is bi-local
and can not be readily used for deriving the Hamiltonian and quantizing it by means of the substitution $g^{\prime}\to -i{\partial/\partial g}$.

In contrast to the Schwarzian action (4) the $\delta g$ fluctuations are scale-invariant and their strength is independent of energy or temperature, being instead controlled by the numerical parameter $N/q^2$. For a finite $q$ the strength of 
such fluctuations decreases with increasing $N$, yet it remains fixed in the double-scaling limit, $N\to\infty$ and $N/q^2=const$.  

Inverting the Hessian operator evaluated at the saddle point (21) requires one to find the    
Green function of the retarded kernel   
%\be
$
D(T,\tau)=<\delta g(T+{\tau\over 2})\delta g(T-{\tau\over 2})>=
<12|K^{-1}|12>
$
which satisfies  the equation 
\bea
\int\int d\tau_5d\tau_6
(-\partial_1\partial_2\delta_{15}\delta_{26}+
<12|{\partial^2 W\over \partial g^2}|56>)
\nonumber\\
<56|K^{-1}|34>
=(\delta_{13}\delta_{24}-\delta_{14}\delta_{23})~~~~~~
\eea
Upon Fourier transforming with respect to the 'center-of-mass' time variable $T$ 
one can use the spectral decomposition  
\bea
D(T,\tau)=\sum_n\int {d\omega\over 2\pi}e^{-i\omega T}
{\psi_{n}({\tau\over 2})\psi^*_{n}(-{\tau\over 2})\over \omega^2-\omega^2_n+i0}\nonumber\\
={i\over 2}\sum_n{e^{i\omega_nT}\over \omega_n}
\psi_{n}({\tau\over 2})\psi^*_{n}(-{\tau\over 2})
\eea
in terms of the eigenfunctions of the equation 
\be
(-\partial^2_{\tau}+{\partial^2 W\over \partial g^2}|_{g_0})
\psi_n(\tau)=\omega^2_n\psi_n(\tau)
\ee
By analogy with the aforementioned averaging over the $\phi$-fluctuations 
the Gaussian average over $\delta g$ 
in the vicinity of the saddle point (22) 
produces the 'Debye-Waller' factor
\bea
{<G^p(\tau)>\over {G^p_0(\tau)}}=<e^{2p\Delta\delta g(\tau)}>=\nonumber\\
\exp( {2p^2\Delta^2} (D(0,\tau)-D(0,0)) )
\eea 
Notably, this averaging is to be performed over the entire function $\delta g$, thereby making no distinction between the 'angular' and 'radial' modes. 

This might be somewhat similar to, e.g., the standard weak-coupling analysis 
of the two-dimensional non-linear $O(N)$ $\sigma$-model 
which seems to emphasize a distinction between the 
longitudinal and transverse fluctuations of the order parameter 
(one gapped and $N-1$ Goldstone  modes, respectively). By contrast,
the exact solution demonstrates no such difference as the true $O(N)$-symmetric 
spectrum consists of the $N$ identical gapped modes. 

Evaluating (20) on the classical trajectory (21) at zero temperature
one finds the effective potential that asymptotically decays at large $\tau$ as $\sim 1/\tau^2$ in both cases of large and small $\gamma$. 
The one-dimensional Green function of the resulting eigenvalue equation   
\be
(-\partial_{\tau}^2+{\kappa\over \tau^2}-\omega^2)\psi=0
\ee
with $\kappa>-1/4$ can be found in the closed form  
\bea
D_{\omega}(\tau,\tau^{\prime})={\pi\over 2i}{\sqrt {\tau\tau^{\prime}}}
{H^{(1)}_{\nu}(\omega\tau_>)\over H^{(1)}_{\nu}(\omega a)}\nonumber\\
(H^{(1)}_{\nu}(\omega\tau)J_{i\nu}(\omega\tau_<)
-H^{(1)}_{\nu}(\omega\tau_<)J_{\nu}(\omega a))
%\nonumber\\\approx {\sin(\Omega({\tau-\tau^{\prime}}))\over \Omega}
\eea
where $\tau_>$ and $\tau_<$ stand for the larger/smaller of $\tau$ and $\tau^{\prime}$, 
respectively, $\nu={\sqrt {1/4+\kappa}}$, and $a$ is the UV cutoff.

For $\omega=0$ (29) amounts to the previously 
derived expression whose finite-temperature version reads \cite{syk01,syk02,syk03,syk04,syk05,syk06,syk07,syk08,syk09,syk10,syk11,syk12,syk13,syk14,syk15,syk16,syk17,syk18,syk19,syk20,syk21,syk22}
\be
D_0(x,x^{\prime})={1\over V\pi}(1+\tan x_{<}({V\pi\over 2}+x_{<}))
(1-\tan x_{>}({V\pi\over 2}-x_{>}))
\ee
where $x=\pi\tau/\beta$ and $V=v+{2\over \pi}\cot \pi v/2$.

Expanding the Bessel functions one once again finds only a mild effect of the Gaussian fluctuations (this time around, of $\delta g$), as the ensuing reduction of the amplitude 
$
{<G^p(\tau)>/G_0^p(\tau)}=\exp(O(1)\Delta^2p^2)
$, 
does not alter the mean-field exponent of the power-law decay.
\\

\section{Quadratic fluctuations in $g$-space}

As an alternative to (26) 
one can formulate the eigenvalue equation in terms of the $g$-variable \cite{JY} 
\be
(-2{{{\sqrt {W_0-W}}}\partial_g{{\sqrt {W_0-W}}}\partial_g}+{\partial^2 W\over \partial g^2}|_{g_0})\psi_n=\omega_n^2\psi_n
\ee
where $W_0=W(g_0)$ without the need  to explicitly 
solve for the classical trajectory $g(\tau)$. 

However, a generally non-trivial derivative $\partial_{\tau}g$
precludes an immediate use of the known solutions such as (11) in the case of, e.g., 
the Morse potential $W(g)$. 
Then treating (31) as a generic second-order equation 
\be
p(x)\partial^2_x\psi+q(x)\partial_x\psi+(E-V)\psi=0
\ee
and eliminating the linear derivative term one 
can convert the equation (31) into the standard Schroedinger equation with the potential
$V^{\prime}=V+{\partial^2_xQ\over Q}$ in terms of the wavefunction $\chi=\psi Q$ with 
$Q(x)=\exp(\int dxq/2p)$. 

Using this equation in the classically accessible 
domain $g_0<g<0$ one can study the system's thermodynamics.
For example, in the case of the Hulten potential (13) one obtains non-trivial
temperature dependences of energy $E=E_0-O(J^{4/3}\beta^{1/3})$ and entropy $S=S_0-O((J\beta)^{4/3})$
which suggest rather peculiar thermodynamic relations. 
\\

\section{Ladder eigenfunctions and chaos exponents}

A chaotic behavior may develop in the complementary (classically prohibited) regime $g<g_0$. 
One popular quantifier of chaos is provided by 
the out-of-time-order correlator (OTOC) given by 
the averaged amplitude $<G_f(\tau_1,\tau_3)G_f(\tau_2,\tau_4)>$ analytically continued from 
the domain $\tau_4<\tau_2<\tau_{3}<\tau_{1}$
to the complex times $\tau_{1}=\beta/4-it/2, \tau_2=-\beta/4-it/2,
\tau_3=it/2$, $\tau_{4}=-\beta/2+it/2$.

On top of a non-exponential regular part of zeroth order in $1/N$ 
the OTOC function demonstrates an exponentially growing first order correction. 
In the case of the $SYK_q-SYK_{q/2}$ model it reads  
\be 
OTOC(t)=O({1\over \beta J})- {f(\gamma)\over N}e^{\lambda_Lt}
\ee
where $f(0)=O(1)$. It is controlled by the Lyapunov chaos exponent determined by the 
ladder eigenstate equation constructed out of the Wightman correlators $G_{lr}(t)=
G(\tau=it+\beta/2)$ \cite{syk01,syk02,syk03,syk04,syk05,syk06,syk07,syk08,syk09,syk10,syk11,syk12,syk13,syk14,syk15,syk16,syk17,syk18,syk19,syk20,syk21,syk22}.
Taking $\omega_n$ in (26) to imaginary values $\omega_n\to i\lambda_L$
yields the exponentially growing ansatz $D(T,t)\sim e^{\lambda_LT}\psi(t)$ 
where the real-time eigenfunction $\psi(t)$ solves the equation \cite{JY} 
\be
(-\partial_x^2-{\cos\theta\over \cosh x+\cos\theta}-
{2\sin^2\theta\over (\cosh x+\cos\theta)^2})\psi=-({\lambda_L\beta\over 2\pi v})^2\psi
\ee
with $\theta=\tan^{-1}(v/J\beta\gamma^2)$. 
Importantly, the potential in (34) is monotonic and
its sign is opposite of that in (26).
  
For $\gamma=0$ this potential is the original SYK's one, $V_0(x)=-2/\cosh^2x$,   
which supports no bound states other than the ground one, 
$\psi_0(x)\sim 1/\cosh x$, with the eigenvalue $E_0=-(\lambda_L\beta/2\pi v)^2=-1$ \cite{syk01,syk02,syk03,syk04,syk05,syk06,syk07,syk08,syk09,syk10,syk11,syk12,syk13,syk14,syk15,syk16,syk17,syk18,syk19,syk20,syk21,syk22}.
As has been repeatedly pointed out in the literature, this value of the chaotic operator growth 
is (almost) maximally possible, its reduction at strong coupling 
($J\beta\gg 1$) being solely due to the temperature-dependent factor $v$
\be 
\lambda_L={2\pi\over \beta}(1-O({1\over \beta J}))
\ee
In the complementary weak 
coupling regime ($J\beta\ll 1$) the chaotic exponent is $\lambda_L\sim J$.

In principle, the rest of the spectrum in (34) could provide for some slower growing terms.
However, for $\gamma=0$ no such terms appear  
as the next (single-node, hence first excited) state would be given by the function
$\psi_1\sim g^{\prime}$ with the eigenvalue $E_1=0$ \cite{JY}.

Also, at longer times $t\gg \beta$ the behavior of the OTOC function is determined by 
the $2$-particle density of states, resulting in another universal power-law, 
$OTOC(t)\propto 1/t^{6}$ \cite{bak01,bak02}. 
 
In Ref.~\cite{lunkin2} a weak $SYK_2$ term in (20) was found not to drastically alter the strong-coupling behavior, except for a reduction of the amplitude by a factor 
$O({1/J\beta\gamma^2})<1$ in the entire interval $1/N\lesssim\gamma\lesssim 1/N^{1/2}$.
At such parameter values the Schwarzian fluctuations were found to be 
suppressed, thus extending the validity of the $SYK_4$ mean-field solution beyond the energy scale $J/N$ all the way down to $J\gamma^2$ at which the Fl behavior finally sets in.
 
In Ref.~\cite{JY}, the chaotic exponent of the large-$q$ bi-quadratic model 
was computed with the use of perturbation theory about the 
state $\psi_0$ for small $\gamma$, thereby finding
\be 
\lambda_L={2\pi\over \beta}(1-O(min[{J\beta\gamma^2},{1\over J\beta\gamma^2}]))
\ee
For comparison, Refs.~\cite{35a,35b} found the exponent  
$\lambda_L={2\pi\over \beta}(1-O({\beta^2\Gamma^2}))$ in the  
$SYK_q-SYK_2$ model, suggesting the possibility
of a finite-temperature transition for arbitrarily small $\Gamma$. 

The latter should, however, be contrasted against the result of Ref.~\cite{cao}  
which reported $\lambda_L\sim 1/J{\beta^2\gamma^3}$ for $\gamma\gg max[1,1/J\beta]$.
Such a behavior conforms to the generic quadratic temperature dependence of $\lambda_L$
in disordered FL and could indicate 
the absence of a genuine finite-temperature phase transition for a 
sufficiently large $\Gamma$.  

Adding to the list of possibilities, 
in Ref.~\cite{JY} some non-maximal (temperature-independent and growing with 
the increasing integer parameter $n$) 
values of $\lambda_L$ were reported on the basis 
of a numerical solution of some other ('variable scaling') model with $W(g)\propto 1/(-g)^n$.

As far as more general potentials $W(g)$ are concerned, the 
Hulten potential (13), for one, falls somewhere in between 
the 'super-symmetric' ($\gamma=1$) point of the $SYK_q-SYK_{q/2}$ model
and the 'variable scaling' one. 
The corresponding eigenvalue equation now reads 
\be
(-\partial_x^2-{2\over {\delta}}
({1\over \cosh x}-{1\over \cosh x+{\delta}}))\psi=-({\lambda_L\beta\over 2\pi v})^2\psi
\ee
where ${\delta}={\sqrt {1+4\gamma^2}}/J\beta\gamma^2$. 
At the super-symmetric point where ${\delta}=5^{1/2}/J\beta$  
and for low temperatures (${\delta}\ll 1$) the potential in the equation
(37) approaches the original SYK's $V_0(x)$ and
the maximally chaotic behavior ($\lambda_L\to 
{2\pi\over \beta v}$) is once again restored. In the opposite limit of 
${\delta}\gg 1$ the potential flattens out
and the Lyapunov exponent decreases monotonically all the way to zero.
It does not vanish at any finite temperature, though, thus calling for a closer look at  
any scenario of a finite-temperature phase transition - or a zero-temperature 
one predicted to occur at a critical $\gamma_c$ vanishing as a power of $1/N$.   
\\

\section{Summary}

This note discussed various generalizations of the SYK model  
that lead to the one-dimensional Liouvillean quantum mechanics. 
Of a particular interest are crossovers 
between the different 
conformal fixed points where all pertinent coupling constants are likely to be of
the same order. Such 'SYK transits' are not directly amenable 
to perturbation theory in the vicinity of the fixed points in question  
but can still be explored in the large-$q$ limit.
To that end, one can utilize the already available -and seek out new - 
non-perturbative mean-field solutions akin to (22) 
which interpolate between the distinct conformal regimes.
This way one could advance the previous studies of 
the bi-quadratic model (20) and its further extensions 
within a broader class of the effective potentials $W(g)$. 

In particular, this preliminary analysis finds that the Lyapunov exponent at the 'super-symmetric' point of the model (20) remains non-zero  
down to the lowest temperatures. This observation may call 
for inspection of the earlier conclusions about the onset of the non-chaotic FL phase
at a critical coupling $\gamma_c$ which could be as weak as $O(1/N^{1/2})$ or even $O(1/N)$ \cite{21a,21b,21c,21d,21e,21f,21g,21h,21i,21j,21k,21l,21m,21n,21o,21p,lunkin2,35a,35b}.  

Also, further generalizations of
the standard Liovillean action related to the various analytically solvable quantum mechanical 
Hamiltonians might be of interest well above and beyond the original SYK context. 

%\end{document}

\end{document}